\documentclass[aps,showpacs,twocolumn]{revtex4}

\usepackage[dvips]{graphicx}
\usepackage{amsmath,array,amssymb} \usepackage{amsfonts} \usepackage{color}

 \newcommand{\s}{\sigma}
\newcommand{\la}{\langle} \newcommand{\ra}{\rangle}
\newcommand{\ua}{\uparrow} \newcommand{\da}{\downarrow}
\newcommand{\be}{\begin{equation*}} \newcommand{\ee}{\end{equation*}}
\newcommand{\bea}{\begin{eqnarray*}}
  \newcommand{\eea}{\end{eqnarray*}} 
  
\newcommand{\G}{\Gamma} 

\begin{document}

\title{The Miscible-Immiscible Quantum Phase Transition in Coupled
  Two-Component \\Bose-Einstein Condensates in 1D Optical Lattices}

\author{Fei Zhan$^{1,2}$, Jacopo Sabbatini$^{1,2}$, Matthew J.
  Davis$^1$, and Ian P. McCulloch$^{1,2}$}

\affiliation{$^1$School of Mathematics and Physics, The University of Queensland, St.~Lucia, QLD 4072, Australia\\
  $^2$Centre of Excellence for Engineered Quantum Systems, The
  University of Queensland, St.~Lucia, QLD 4072, Australia}

\begin{abstract}
  We study the miscible-immiscible quantum
  phase transition in a linearly coupled binary Bose-Hubbard model
  in one dimension that can describe the low-energy properties of a
  two-component Bose-Einstein condensate in optical lattices.  With
  the quantum many-body ground state obtained from density matrix
  renormalization group algorithm, we calculate the characteristic
  physical quantities of the phase transition controlled by the linear
  coupling between two components. Furthermore we calculate the Binder
  cumulant to determine the critical point and construct the phase diagram.
  The strong-coupling expansion shows that in the Mott insulator
  regime the model Hamiltonian can be mapped to a spin $1/2$ XXZ model
  with a transverse magnetic field.
\end{abstract}
\pacs{64.70.Tg, 
  03.75.Nt, 
  03.75.Mn, 
  05.10.Cc} 
\date{\today}

\maketitle

\section{Introduction}
\label{sec:introduction}
In recent years, the progress in single-atom detection and the manipulation
of ultra-cold neutral atoms has allowed experimentalists to use
these systems to engineer and emulate condensed matter systems
\cite{greiner2002n,bloch2012np}.  Multi-component Bose-Einstein
condensates (BEC), formed by atoms of different atomic species or
different hyperfine states, has attracted attention from both
experimentalists~\cite{stenger1998n,hall1998prl} and theorists
\cite{hotinlun1996prl,puh1998prl,timmermans1998prl} due to their
larger symmetry groups and extensive degrees of freedom. As the
simplest model consisting of multiple components, the binary BEC has
been an appealing candidate to simulate the spin-$1/2$ fermionic
superconductor, magnetic behavior~\cite{paredes2003prl}, superfluids
\cite{kuklov2004prl}, phase separation~\cite{alon2006prl}, quantum
phase transitions~\cite{sabbatini2011prl,sabbatini2012njp} and
thermalization~\cite{zhangjiangmin2012pra}.

Binary BECs are naturally divided into miscible and immiscible
mixtures based on the interaction strength characterizing the system.
In a two-component BEC we can observe three kinds of
interactions: interaction within the first component,
interaction within the second component,
and interaction between the two components.
If the strength of the inter-component
interaction exceeds that of the intra-component interaction, then energy
considerations show that the two-components prefer to be in a phase
separated or immiscible state~\cite{alon2006prl,altman2003njp}. If the opposite
is true then the system is said to be in the miscible phase.

A two-component BEC composed of bosons in different hyperfine states
can, however, undergo a miscible-immiscible phase transition controlled
by a linear coupling between the energy levels \cite{merhasin2005jpbamop}.
This phenomenon has been studied in a number of settings, such as
nonlinear Josephson-type oscillations
\cite{williams1999pra}, non-topological vortices
\cite{parkqhan2004pra} and non-equilibrium dynamics across the
critical point~\cite{leechaohong2009prl,nicklas2011prl,de2014pra}.

The dependence of the order parameter on the linear coupling
coefficient revealed a second-order transition in a mean-field (MF)
numerical study of this phase transition~\cite{merhasin2005jpbamop}.
The properties of a second-order phase transition depend entirely on
its universality class and are insensitive to the microscopic details
of the underlying system. The universality class is determined by a set
of power law indices, called critical exponents, which characterize
quantities such as the correlation length and the response time of the
system~\cite{cardy1996}.  Studies of the static properties of a system
near the critical point are challenging because of the divergence of
these quantities.

In this paper we study the miscible-immiscible phase transition of a
linearly coupled two-component Bose-Hubbard model describing the low
energy physics of a binary BEC loaded in an optical lattice in one dimension.

The exponentially growing size of the Hilbert space as the lattice
grows in size prevents the investigation of the full quantum state
with exact diagonalization methods, even for lattices of moderate
sizes.  On the other hand, matrix product states (MPS) can
parameterize the size of the Hilbert subspace relevant to the low
energy properties by the dimension of the matrices, {\it i.e.}, the
number of states, and the size of the subspace grows polynomially with
the lattice size~\cite{schollwock2011ap}.  In this paper we employ MPS as the ansatz to
represent the many-body state and density matrix renormalization group
(DMRG) to variationally solve for the ground-state. Infinite DMRG (iDMRG)
methods~\cite{McCulloch08} exploiting the system's translational invariance in the
thermodynamic limit allow us to compute the ground state of the
system without boundary or finite size effects.  From the ground state
we can determine a variety of observables like expectation values and
multi-point correlations that help us characterize the quantum phase
transition and its critical exponents. With the iDMRG methods the
correlation length can be calculated directly from the
eigenvalues of the transfer matrix.

The paper is organized as following: In Section \ref{sec:model},
we describe the model Hamiltonian for the numerical calculation and the
definitions of order parameter and correlation function.  In Section
\ref{sec:finite}, we present our results for the mean occupation number
distribution, correlation function, correlation length, phase diagram,
and entanglement entropy obtained with a finite system.  The
calculations for the infinite system are described
in Section \ref{sec:idmrg}.  In Section \ref{sec:u1z2}, we illustrate
the ground state in a different set of basis states that are
categorized by the $\mathbb{Z}_{2}$ symmetry. Finally we conclude in Section
\ref{sec:conclusion}.

\section{Model Hamiltonian and symmetries}\label{sec:model}
The system we consider here is a binary BEC in a 1D
optical lattice with lattice constant $L_{0}$.
The length scale is chosen such that $L_{0}=1$.
The binary BEC consists of two hyperfine atomic states of a
single species, which can be defined as spin-up and spin-down,
$\s=\uparrow,\downarrow$.  Spins of two different orientations are
coupled by a two-photon transition.  This can be realized in an
ultra-cold atom gas experiment with e.g. $^{87}$Rb atoms
\cite{hall1998prl}.

The binary BEC in an optical lattice can be mapped to a two-component
Bose-Hubbard model, which is composed of three parts,
\begin{equation}
  \hat{H} = \hat{H}_{0} + \hat{H}_{I} + \hat{H}_{C},\label{eq:hamiltonian}
\end{equation}
where the three portions of the total Hamiltonian are given by
\begin{eqnarray}
  \hat{H}_{0} & = & -J\sum_{j=1;\s}^{L-1}\left[a_{j+1,\s}^{\dagger}a_{j,\s}+H.c.\right],\label{eq:h0}\\
  \hat{H}_{I} & = & \frac{U}{2}\sum_{j=1;\s}^{L}n_{j,\s}(n_{j,\s}-1)+U_{\ua\da}\sum_{j=1}^{L}n_{j,\ua}n_{j,\da}, \label{eq:hi}\\
  \hat{H}_{C} & = & -\Omega\sum_{j=1}^{L}\left(a_{j,\ua}^{\dagger}a_{j,\da}+a_{j,\da}^{\dagger}a_{j,\ua}\right),\label{eq:hc}
\end{eqnarray}
respectively. In the above Hamiltonian, $a_{j,\s}^{\dagger}(a_{j,\s})$
creates (annihilates) a boson with spin orientation $\s$ on the $j$th
site and $n_{j,\s}=a_{j,\s}^{\dagger}a_{j,\s}$ is the corresponding occupation
number operator.  Bosons of either spin species can
tunnel to the nearest-neighbor site with tunneling energy
$J$, assumed here to be the same for both species.  
Only on-site interactions are included, with interaction energy
$U$ between same-spin species, and $U_{\uparrow\downarrow}$ between
different spins.  The amplitude of the two-photon microwave coupling
between two components is denoted as $\Omega$.  In our calculation we
define the energy unit such that such that $J=1$.

The interplay of the intra-component interaction and
inter-component interaction determines the phase of the binary BEC
\cite{hotinlun1996prl}. With no inter-component coupling, $\Omega=0$, 
and large intra-component interaction $U> U_{\ua\da}$,
the total energy is minimized by spreading each
components equally to all sites. On the other hand, whenever
$U_{\ua\da} > U$ the system phase-separates~\cite{Comment}.
This distinction can be quantified as:
\begin{equation}
  \Delta=\frac{U^{2}}{U_{\ua\da}^{2}}
\end{equation}
where $\Delta>1$ indicates a miscible phase and $\Delta<1$ is phase-separated.

Turning on the inter-component coupling, $\Omega>0$, the symmetry is reduced from $U(1)\times U(1)$ to
$\mathbb{Z}_{2} \times U(1)$ and the phase-separated state is replaced by an immiscible phase,
analogous to a spin ferromagnet, where the occupation number of each component differs
but the system remains translationally invariant. For sufficiently large $\Omega$, the
system is always in the miscible phase. In this paper, we consider the case
$\Delta=1/4$ and tune $\Omega$ to realize the
miscible-immiscible phase transition.

In Section \ref{sec:finite}, we consider open boundary conditions
(OBC) for a finite system of $L$ sites and $N$ total number of
particles.  Even though OBC brings forth obstructive boundary effects,
it is numerically less expensive than periodic boundary conditions.  
Indeed, the DMRG algorithm with periodic boundary condition demands additional 
efforts for an effective simulation~\cite{pippan2010prb}.

We will focus on the phase transitions with a global filling factor
$\rho=N/L=1$.  Such a system in an optical lattice can be
appropriately mapped to a single-band Bose-Hubbard model
\cite{oosten2003pra,greiner2001prl}.

To look into the miscible-immiscible transition we will study the
expectation value and correlation function of the occupation number
difference operator, which on the $j$th site is defined as
$\Delta N_{j} = n_{j,\uparrow}-n_{j,\downarrow}$.
Its average throughout the whole lattice 
\begin{equation}
  M=\frac{1}{L}\sum_{j}^{L} \Delta N_{j}  \label{eq:orderparameter} \;,
\end{equation}
is the order parameter of the phase transition in this model.
The expectation value of this order parameter is the magnetization of the system.
Note that neither this order parameter, nor the particle number operator 
for each component, commutes with the total Hamiltonian
\eqref{eq:hamiltonian}, due to the coupling of Eq. \eqref{eq:hc}.

We also study the correlation function of occupation difference
operators between bosons on the $j$th site and $j'$th site
\begin{equation}
  C(j,j')=\la\Delta N_{j}\Delta N_{j'}\ra.
  \label{eq:correlation}
\end{equation}
If the system is translationally invariant, $C(j,j')$ only depends on
the distance between the two sites $|j-j'|$, and thus we can define
$C(j)=C(0,j)$.

It is important to consider the symmetries of the model.
First, the Hamiltonian has $U(1)$ symmetry as it commutes with the total number operator $N$.
Second, the Hamiltonian has a discrete $\mathbb{Z}_{2}$ symmetry as the
Hamiltonian is unchanged if all the spins are flipped.  We will
show in the following sections that this symmetry is
spontaneously broken when $\Omega$ is below the critical value.  

\section{Finite system results}
\label{sec:finite}

In this paper we consider the phase transition occurring in Mott
insulator regime.  Actually the critical point separating the SF and MI
regime has not been documented in the literature for a linearly
coupled two-component BEC in optical lattices~\cite{Note1}.
In the Mott
insulator regime the energy scale of the system is dominated by the
on-site interaction energy.  The coexistence of multiple bosons on a
same lattice site is energetically expensive for integer filling and
thus particles are equally spread over all lattice sites.
The local particle number fluctuation vanishes in the ground state.
The excited state is gapped from the ground state in the Mott
insulator regime and contains pairs of quasi-particles and quasi-holes.

MPS is an excellent ansatz for the ground state of a gapped system.
A mean-field derivation can approximate this system in the superfluid
regime and predicts the critical value for $\Omega$ as
\cite{sabbatini2011prl,sabbatini2012njp},
\begin{equation}
  \Omega_{\rm c}=U\rho\left(\frac{1}{\sqrt{\Delta}}-1\right).
\end{equation}
Below we show that MPS provides a more accurate MI
ground state than the mean-field theory, and predicts a
different power law dependence of $\Omega_{c}$ on $U$ that agrees
with a second-order perturbation theory.

A basic question is still open: where is the border between
superfluid regime and Mott insulator regime for this linearly-coupled
two-component Bose-Hubbard model?  For this two-component model, more
degrees of freedom give rise to two branches of the quasiparticle spectrum
\cite{tommasini2003pra}.  In the absence of the coupling $\Omega$ between the
two components the two branches are independent.  When the coupling is
turned on, one branch accounts for SF-MI transition but the
other depends on $\Omega$ and is responsible for miscible-immiscible
phase transition, which will be addressed in Section \ref{sec:u1z2}.

The border can be determined by locating the value of on-site
interaction where the energy gap closes.  The energy gap can be simply
verified numerically by calculating the ground state energy $E$ for
systems of $N-1$, $N$, and $N+1$ total particles.  The system has
integer filling factor when the total number of particles is $N$.  The
energy gap $\Delta E$ can be obtained by the formula,
\begin{equation}
  \Delta E = E(N+1) + E(N-1) - 2E(N).\label{eq:sfgap}
\end{equation}
Within numerical accuracy, the simulation gives $\Delta
E\neq 0$ for a system with parameters in MI regime.  This will be
addressed in a subsequent publication~\cite{InPreparation}.

In the following finite DMRG calculations, we choose the number of
states $m=300$ for the MPS, which is large enough to ensure the variational ground state is close
to the true ground state, while being computationally efficient.

\begin{figure}[t]
  \includegraphics[width=\linewidth]{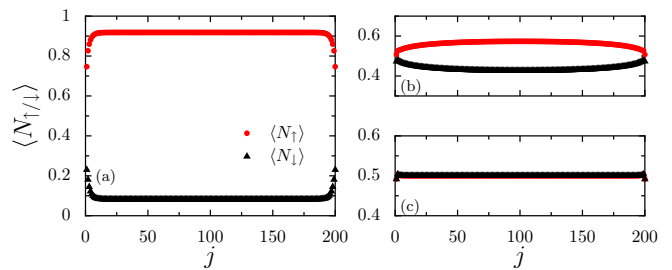}
  \caption{\label{fig:mi_occ}(Color online) Mean occupation number for
    spin-up bosons $\langle N_{\ua}\rangle$ (red circles) and
    spin-down bosons $\langle N_{\da}\rangle$ (black triangles) on the
    $j$th site throughout a $200$-site lattice with open boundary
    condition (a) below, (b) near but still below, and (c) above the
    critical point in Mott insulator regime. The values for the interactions are $U=5, U_{\ua\da}=2U$.
  The values for the linear coupling are (a) $\Omega=0.18$, (b) $\Omega=0.214$, and (c) $\Omega=0.248$.}
\end{figure}

\subsection{Occupation distribution}
\label{subsec:occupation}
We first show the immiscible and miscible phases by displaying the
occupation distribution throughout a lattice.  In Fig.~\ref{fig:mi_occ},
we plot the mean occupation number for
(a) $\Omega\ll\Omega_{c}$, (b) $\Omega\lesssim\Omega_{c}$, and (c) $\Omega\gg\Omega_{c}$.

When $\Omega\ll\Omega_{c}$, the system is in the immiscible phase,
where the largest energy scale in the system is the
inter-component interaction strength.  The coexistence of different
boson species costs more energy compared to the same species,
therefore states with only one
component on each sites are favorable.
We recall the total Hamiltonian \eqref{eq:hamiltonian} preserves the
$\mathbb{Z}_{2}$ symmetry, as the energy is unchanged when all spin
orientations are flipped.
In the thermodynamic limit, the ground state is 2-fold degenerate.
The $\mathbb{Z}_{2}$ spontaneous symmetry breaking will occur in
this regime of $\Omega$.
On the other hand, in principle for a finite
size system spontaneous symmetry breaking should not occur.
Nevertheless, obviously in Fig.~\ref{fig:mi_occ}(a) the ground state in
immiscible phase does not preserve the $\mathbb{Z}_{2}$ symmetry of
the total Hamiltonian with the imbalance $\la\Delta N\ra\neq0$ in mean
occupation numbers for two components.

In the ground state subspace, the DMRG variational calculation adopts 
the lowest-entropy state and therefore numerically enforces the 
order of the symmetry breaking state. 
In a numerical DMRG calculation the broken symmetry state
is variationally favored if the energy splitting of the ground state
is smaller than the energy scale set by the truncation error of the calculation.
There is randomness in this favoredness.
In a real-life numerical simulation many factors, {\it e.g.}, the direction of
DMRG variational algorithm, may determine which component will appear
in the favorable states.  In order to show the randomness, for each
values of $\Omega$ we start the DMRG simulation with a different
random initial wave function.  The probability that one of the two
components is preferred by random fluctuation is one half.
Consequently, when the system is in the immiscible phase we saw the
domination of spin-up bosons in half of the simulations and the
domination of spin-down bosons in the other half (not shown here).

As the coupling coefficient $\Omega$ increases, the imbalance in
occupation decreases, reaching zero at critical point as can be seen in Fig.
\ref{fig:mi_occ}(b).  Above the critical point, the ground state
has the same $\mathbb{Z}_{2}$ symmetry of the Hamiltonian and
both components equally occupy all of the sites.  Therefore the
imbalance must be zero and the system is in the miscible
phase, see Fig.~\ref{fig:mi_occ}(c).

One must always be aware of the boundary effects when we approximate a system
in thermodynamic limit with a finite system. The boundary effect comes
from the correlation between a particle in the bulk of the finite
system and a particle on the boundary, where particles can only hop in
one direction.  As we can see in Fig.~\ref{fig:mi_occ}(a), near the
boundary the mean occupation number for both spin-up and spin-down
bosons deviates from the bulk.  It decreases
partially the mean occupation number of the dominant component, and
increases that of the other component.  Approaching the critical point,
the correlation between two sites at longer distance starts to become
non-negligible, as we expect for a second-order phase transition.
Consequently the boundary effect more strongly influences the sites
in the bulk of the lattice, as can be seen in Fig.
\ref{fig:mi_occ}(b).  When the coupling coefficient is sufficiently close
to the critical point, the influence of both boundaries merge
together and we see two curved lines for the mean occupation number of
both components.  The boundary effect is negligible
above the critical point, see Fig.~\ref{fig:mi_occ}(c).  The
effect of the boundaries on the calculation of correlation function is
explored in more detail in the following subsection.

\begin{figure}[t]
  \includegraphics[width=.9\linewidth]{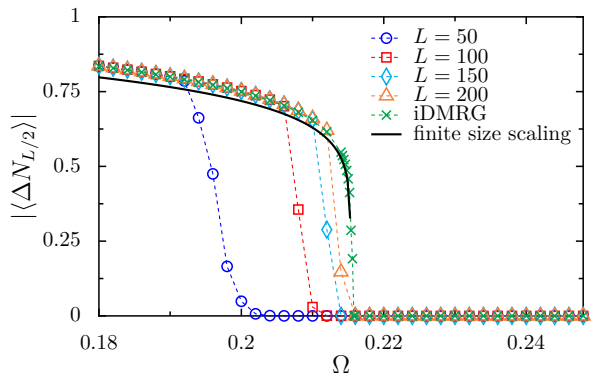}
  \caption{\label{fig:mi_occ_fi}(Color online) The absolute occupation
    imbalance $|\la\Delta N\ra|$ on the middle site $L/2$ of lattices
    having $50$ (blue circle), $100$ (red square), $150$ (cyan
    diamond), and $200$ (orange triangle) sites as the function of the
    coupling coefficient $\Omega$, when $U=5, U_{\ua\da}=2U$.
    The green crosses are the data for
    an infinite lattice from iDMRG calculation.  The dashed
    lines are used to guide eyes.  The solid black line is drawn with
    parameters from a finite size scaling about the critical point
    with exponent $\beta=1/8$. }
\end{figure}

In Fig.~\ref{fig:mi_occ_fi}, we show the
absolute imbalance $\la\Delta N_{L/2}\ra$ as a function of the
coupling coefficient $\Omega$ for different lattice sizes. As we
have discussed above, below the critical value $\Omega_{c}$ the
imbalance is non-zero and drops quickly to zero at the critical point. 
We can also see $\Omega_{c}$ shows clear saturating
behavior as the lattice size is increased,
down and $\Omega_{c}$ saturates asymptotically in the thermodynamic
limit.  From Fig.~\ref{fig:mi_occ_fi}, the critical point can be
estimated as $\Omega_{c}\approx0.215$. We improve on this estimate
in section \ref{sec:Binder} below.

By using finite sizing scaling, we can extract the critical coupling
$\Omega_{c}$ and critical exponents by collapsing curves for lattices
of different lengths~\cite{barber1983}.  First we define the reduced coupling
$\epsilon=|1-\Omega/\Omega_{c}|$.  Previous studies in the literature
have indicated this phase transition is of second order
\cite{merhasin2005jpbamop}.  Near the critical point of a second-order
phase transition, we know that the correlation length and
magnetization satisfy $\xi\propto\epsilon^{-\nu}$ and
$M\propto\epsilon^{\beta}$ (only below the critical point, otherwise
$M=0$), from which we can deduce the relation
$M\propto\xi^{-\beta/\nu}$.  For a finite lattice, instead of
approaching zero when $\xi$ diverges, $M$ stays at a finite nonzero
value when $\xi$ becomes comparable to the lattice length $L$.  This
behavior can be described by $M=\xi^{-\beta/\nu}M_{0}(L/\xi)$ with the
assisting function $M_{0}(x)$ that goes to zero as $x^{-\beta/\nu}$
when $x\rightarrow0$ and a constant when $x\rightarrow\infty$.  To
remove the size dependence, we define the scaling function
\begin{equation}
  \tilde{M}(L^{1/\nu}\epsilon)=L^{\beta/\nu}M(\epsilon).
\end{equation}
With $\nu=1$ and $\Omega_{c}=0.2153$ we will obtain in the next
subsections, we find four curves for the four lengths coalesce with
$\beta=1/8$.  In Fig.~\ref{fig:mi_occ_fi}, we plot the curve
$\epsilon^{\beta}$ with $\beta=1/8$ obtained from the finite size
scaling and $\Omega_{c}=0.2153$.  We find near the critical point it
agrees very well with the results from iDMRG calculations in Section
\ref{sec:idmrg}.

\subsection{Correlation function and correlation length}
\label{subsec:cfcl}
\begin{figure}[t]
  \includegraphics[width=.9\linewidth]{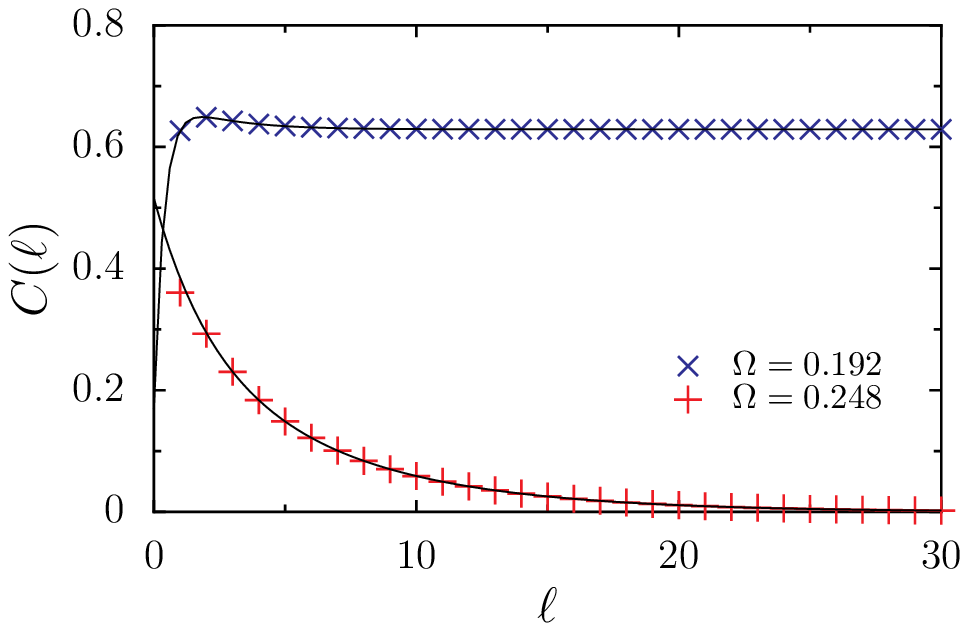}
  \caption{\label{fig:mi_corr}(Colour online) The correlation function
    $C(\ell)$ calculated in the bulk of a lattice with $200$ sites
    with $60$ sites cut off at both ends.  The blue crosses (red
    pluses) are the values of the correlation function between the
    zeroth site and the $\ell$th site when $\Omega=0.192$
    ($\Omega=0.248$), which is below (above) the critical point.  The
    curves are the fittings with respect to a sum of two
    exponential functions. The values for interactions are $U=5, U_{\ua\da}=2U$.}
\end{figure}

\begin{figure}[t]
  \includegraphics[width=.9\linewidth]{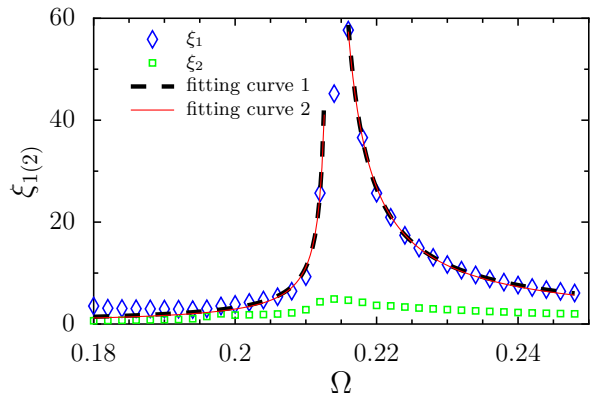}
  \caption{\label{fig:mi_corr_p200}(Colour online) The longer (blue
    diamond) and shorter (green square) longest correlation lengths
    extracted by fitting the correlation function with a sum of
    two exponential functions for a lattice of $200$ sites.  The
    longer correlation length shows divergence in a certain region of the
    coupling coefficient $\Omega$.  The black dashed line shows the
    fitting curve with the critical exponent $\nu$ as an independent
    variable and the red solid line shows the fitting curve with the
    plausible fixed exponent $\nu=1$, see text. The values for interactions are $U=5, U_{\ua\da}=2U$.}
\end{figure}

Close to the boundary the quantum state may deviate
dramatically from the infinite system due to
the inevitable boundary effect (see Fig. \ref{fig:mi_occ}).
For this reason, in order to simulate the
correlation in the thermodynamic limit with a finite size system, we
only investigate the correlation function computed in the bulk of the
lattice, where boundary effect is minimized.  Keeping this in mind
is particularly important when $\Omega<\Omega_{c}$, see Fig.
\ref{fig:mi_occ}(a).  Very close to critical point the boundary effects cannot be neglected.
In this region the correlation function is not reliable
for extracting the correlation length.
For instance, to calculate the two-point correlator
\eqref{eq:correlation} in Fig.~\ref{fig:mi_corr} for a lattice of
$200$ sites, we cut off $60$ sites at both ends of the lattice.
In Fig.~\ref{fig:mi_corr} we show the correlation function up to $30$ sites.
The correlation function only depends on the distance $|j-j'|$ between its two studied sites.
Consequently we only enumerate the distance by $\ell=|j-j'|$.
For $\Omega=0.192\ll\Omega_{c}$, in Fig.~\ref{fig:mi_corr}, we find at short distances
the correlation function first increases and then
decreases exponentially before saturating at a nonzero value.
The short range most likely stems from the finite-size effects,
since in iDMRG calculations the correlation function only
decreases exponentially and saturates to a nonzero value.
For $\Omega=0.248\gg\Omega_{c}$ in the miscible phase,
the correlation function exponentially decays to zero.

As suggested by the characteristic form of correlation functions
in MPS~\cite{ostlund1995prl,schollwock2011ap} and
the fact that correlation decays exponentially in a system away from
criticality, the correlation function can be fitted with a sum of exponential functions.
Here we fit the correlation function with a sum of two exponential functions:
\begin{equation}
  C(\ell)=\sum_{i=1,2}a_{i}\exp(-\ell/\xi_{i})+c\label{eq:fittingfunction},
\end{equation}
where the constant $c$ has a nonzero value when $\Omega$ is below the
critical point.  We find in Fig.~\ref{fig:mi_corr} that the fitting
precisely captures the behavior of the correlation function.

In Fig.~\ref{fig:mi_corr_p200}, we plot the two correlation lengths from the fitting function
\eqref{eq:fittingfunction}. The longer correlation length $\xi_{1}$ shows clear
divergent behavior around $\Omega=0.215$, which characterizes the behavior of
the system close to criticality. In principle, the shorter correlation
length $\xi_{2}$ will also diverge at the critical point~\cite{cardy1996}, but this
is difficult to fit from finite size data because the correlation length
is much shorter than $\xi_{1}$.
We fit the correlation length with the
power law $\xi_{1}\propto|\Omega-\Omega_{c}|^{-\nu}$ around critical
point.  In the first fitting, shown in Fig.~\ref{fig:mi_corr_p200}, we
set $\nu$ as an independent variable and obtain the following
optimized fitting function:
\begin{displaymath}
  \xi_{1}(\Omega) = \left\{
    \begin{array}{lr}
      \frac{0.07144\pm0.0342}{|0.2133\pm0.0001-\Omega|^{0.8844\pm0.0336}} & : \Omega<\Omega_{c}\\
      \frac{0.2797\pm0.0369}{|0.2131\pm0.0002-\Omega|^{0.915\pm0.0336}} & : \Omega>\Omega_{c}.
    \end{array}
  \right.
\end{displaymath} 
On account of the nonlinear least square algorithm's numerical
complexity, combined with the less reliable data near the
critical point, the exponent $\nu$ may have low numerical accuracy.
Nevertheless, they are close to the already known value $\nu=1$
in the conformal field theory (CFT) for the 1D quantum Ising model with a
transverse magnetic field.  Therefore for a second fitting, also shown
in Fig.~\ref{fig:mi_corr_p200}, we set $\nu=1$ and
obtain:
\begin{displaymath}
  \xi_{1}(\Omega) = \left\{
    \begin{array}{lr}
      \frac{0.04014\pm0.00825}{|0.2135\pm0.0004-\Omega|} & : \Omega<\Omega_{c}\\
      \frac{0.2014\pm0.005}{|0.2125\pm0.0001-\Omega|} & : \Omega>\Omega_{c}.
    \end{array}
  \right.
\end{displaymath}
The closeness of these two fit functions in Fig. \ref{fig:mi_corr_p200}
shows the difficulty in obtaining accurate values of $\nu$ and $\Omega_{c}$
by this method. The critical point $\Omega_{c}=0.213$ is somewhat
below the result obtained from Fig.~\ref{fig:mi_occ_fi} ($\Omega_{c}=0.2153$), and demonstrates the
accuracy of this fitting technique.
The fitting for the parameter above the critical point
is better than the other side in the coupling coefficient space.  The
reason is due to the more severe boundary effect below the critical
point.  It can be seen in Fig.~\ref{fig:mi_corr_p200} that both
curves fit the data points very well, with only slight
deviations when the coupling coefficient is far away from the critical
point.  In any event, this suggests that $\nu=1$ is likely,
as consistent with already known theories.

\subsection{Binder cumulant}
\label{sec:Binder}
\begin{figure}[t]
  \includegraphics[width=.9\linewidth]{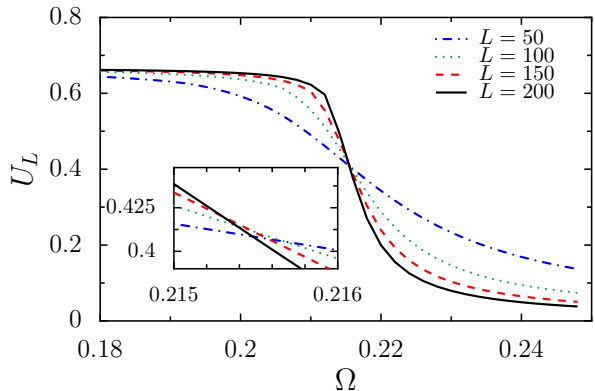}
  \caption{\label{fig:mi_binder} (Color online) The Binder cumulant $U_{L}$ Eq. \eqref{eq:binder} as
    a function of the coupling coefficient $\Omega$ for lattices of
    $L=50$ (blue dash-dotted), $100$ (green dotted), $150$ (red
    dashed), and $200$ (black solid), when $U=5, U_{\ua\da}=2U$.  The inset zooms into the region
    where four curves cross each other, near $\Omega_{c}=0.2153$.}
\end{figure}

The Binder cumulant can be used to more accurately determine the critical point
in the thermodynamic limit from finite size calculations~\cite{binder1981prl,kastening2013pre}.
Its potential usefulness and generalizations still attract a lot of
attention~\cite{hasenbusch2008jsmte,selke2009pre}.  With the
Binder cumulant the critical point can be determined with a relatively
small finite size lattice.  For example, the critical temperature for
a two-dimensional Ising model can be obtained from the Binder cumulant with
a $9\times9$ lattice~\cite{binder1981prl}.

The Binder cumulant for this system $U_{L}$ is defined as
\begin{equation} \label{eq:binder}
  U_{L}=1-\frac{\la M^4\ra}{3\la M^2\ra^2},
\end{equation}
where $\la M^2\ra$ and $\la M^4\ra$ are the second-order and the
fourth-order moments of the order parameter, respectively.  Note that
Binder cumulant depends on the length $L$ of the lattice.

In Fig.~\ref{fig:mi_binder}, we plot the Binder cumulant with the same
parameters and the same OBC we have used for the preceding subsections for lattices of a
variety of lengths.  The curves clearly show the asymptotic behavior
of Binder cumulant: It decreases with increasing $\Omega$ and
asymptotically approaches to $2/3$ and $0$ below and above the
critical point, respectively. Near the critical point, it decreases
faster than in other regions. In addition, the data for a larger lattice
exhibits a steeper transition near $\Omega_{c}$.  As a result, the
different curves cross each other at the critical point.  In
the thermodynamic limit the curve should be discontinuous at the
critical point.
 
As we can see in Fig.~\ref{fig:mi_binder},
the four curves cross in a small range of $\Omega$.
The inset of Fig.~\ref{fig:mi_binder} shows the
crossing is located in a region $[0.2152\;\;0.2158]$.  The value of
$\Omega$ for the crossing point corresponds to the critical value
$\Omega_{c}$.  The cubic spline interpolation of the curves for
$L=150$ and $200$ suggests the critical point should be at
$\Omega_{c}=0.2153$.  In Section \ref{subsec:occupation}, we
performed the finite size scaling with this value as the tentative
critical value and obtained the expected value for the exponent $\beta$.
In the following section we will see this value also agrees with the iDMRG results.

\subsection{Phase diagram}
\begin{figure}[t]
  \includegraphics[width=.9\linewidth]{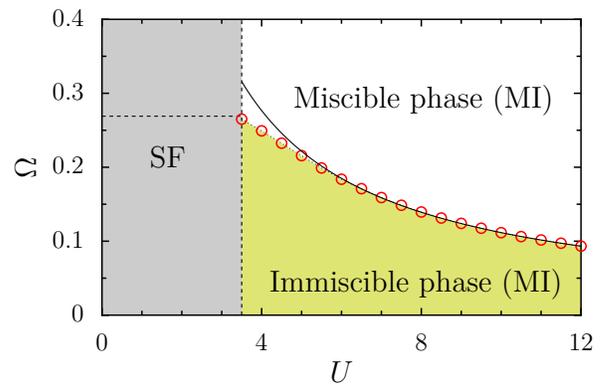}
  \caption{\label{fig:ph_diag} (Color online) The phase diagram in the
    space of coupling parameter $\Omega$ and on-site intra-component
    interaction $U$.  The inter-component interaction $U_{\ua\da}=2U$.
    The border between superfluid and Mott insulator will be updated
    in upcoming works.  The red circles are the data points of the
    border between miscible and immiscible phases determined by Binder
    cumulant for lattices of $L=50$ and $L=100$.  The dotted curve
    connecting the red circles interpolates the data points
    (red circles).  The solid curve is attained by fitting the data
    points with $C/U$ for $U>10$.}
\end{figure}

As seen in the last subsection, the Binder cumulant can locate the
critical point very precisely. Using this measure,
in Fig.~\ref{fig:ph_diag}, we now plot the
phase diagram of this model Hamiltonian in the space of
$\Omega$ and $U$, while keeping $U_{\ua\da}=2U$, corresponding to $\Delta=1/4$.  
In Fig.~\ref{fig:mi_binder},
we see that increasing the number of lattice sites only changes the
value of $\Omega_{c}$ in the fourth digit after the decimal point.
As a consequence, to speed the calculation, we locate $\Omega_{c}$ by using
Binder cumulants for shorter lattices of $L=50$ and $L=100$.

In Fig.~\ref{fig:ph_diag} for the MI regime, we see the critical
value $\Omega_{c}$ decreases as $U$ is increased as approximately $1/U$.
This result is in contrast to mean-field predictions, which shows linear dependence of
$\Omega_{c}$ on $U$~\cite{sabbatini2012njp}. 

When $U$ and $U_{\ua\da}$ are large, the tunneling between sites
is negligible.
Perturbation theory can be employed in the parameter $J$ to find a further approximation
to provide more insight into the underlying physics.  Using the
strong-coupling expansion, we derived the effective Hamiltonian for
this model, which turns out to be a ferromagnetic XXZ model with a
transverse magnetic field:
\begin{align}
  H=&-J_{\bot}\sum_{j}[S^{x}_{j}S^{x}_{j+1}+S^{y}_{j}S^{y}_{j+1}]+J_{z}\sum_{j}S^{z}_{j}S^{z}_{j+1}\notag\\
  &+\Gamma\sum_{j}S^{x}_{j},\label{eq:xxzhamiltonian}
\end{align}
where $S^{x}_{j}$, $S^{y}_{j}$, and $S^{z}_{j}$ are the three
components of the spin-operator for a spin $1/2$ particle on the $j$th
lattice site, respectively.

The coefficients for the effective Hamiltonian are
\begin{align}
  J_{\bot}=&\frac{4}{U_{\ua\da}}, \;\;J_{z}=\frac{4}{U_{\ua\da}}-\frac{8}{U},\notag\\
  \Gamma=&-2\Omega.
\end{align}
For the parameters we have chosen $U_{\ua\da}=2U$, {\it viz.}
$|J_{z}/J_{\bot}|=3$, which indicates the ferromagnetic ground state
when $\Omega=0$.  When $|J_{z}/J_{\bot}|\rightarrow\infty$, {\it i.e.},
the first term in Eq.~\eqref{eq:xxzhamiltonian} can be neglected,
this model can be further mapped onto the Ising model in a transverse magnetic field,
for which we know the phase transition occurs at $\Gamma_c=J_{z}/2$.
For non-zero but small $J_{\bot}$, the exact dependence of $\Gamma_c$ on $J_\bot$ and $J_z$
is not known, but we expect that $\Gamma_c \propto 1 / U$.

In Fig.~\ref{fig:ph_diag}, in the region where $U>10$ we fit the data
points on the border line between miscible and immiscible phase to the
function $C/U$ and find the coefficient $C\approx 1.09$, in excellent
agreement with what we obtained from the DMRG calculation for XXZ model with a transverse field,
where we found $\G_{c}=0.36J_{z}$. Since $J_{z} = -6/U$, we obtain $\Gamma_{c} = 2.16/U$  and $\Omega_{c} = 1.08/U$.

\subsection{Entanglement entropy}

It has been demonstrated that the entanglement entropy,
which is a significant concept in quantum information, also plays an
important role in understanding quantum phase transitions in
condensed matter physics since it is
related to the appearance of long-range correlations
\cite{osborne2002pra,vidal2003prl}.  The bipartite
entanglement entropy can capture the large-scale behavior of quantum
correlations in the critical regime.  In the vicinity of the critical
point $\Omega_{c}$, the entanglement entropy diverges logarithmically with the
correlation length.  Here we will present the scaling behavior of
entanglement entropy for the model and determine the central charge at criticality.

Suppose the lattice is divided into the sublattice $A$ on the left and
the sublattice $B$ on the right.  We define the entanglement entropy as the
von Neumann entropy of either one of the two sublattices, say the
sublattice $A$,

\begin{equation}
  S=-{\rm Tr}(\rho_{A}\log(\rho_{A})),
\end{equation}
where $\rho_{A}={\rm Tr}_{B}(\rho)$ is the reduced density matrix for
the part $A$.

\begin{figure}[t]
  \includegraphics[width=.9\linewidth]{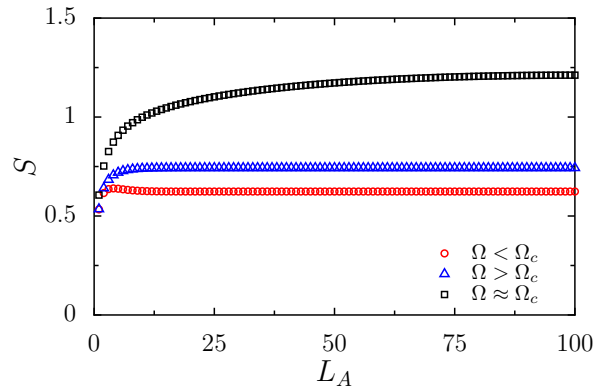}
  \caption{\label{fig:mi_enp} (Color online) The entanglement entropies
    up to half of a $200$ site lattice, below (red circles), near (blue triangles), and above (black squares)
    the critical point in Mott insulator regime. The interaction strengths are $U=5, U_{\ua\da}=2U$.}
\end{figure}

In Fig.~\ref{fig:mi_enp}, we plot the entanglement entropy for
$\Omega<\Omega_{c}$, $\Omega\approx\Omega_{c}$, and
$\Omega>\Omega_{c}$ as the size of the sublattice $A$ is increased up
to the half of the whole lattice.  We can find the bipartite
entanglement entropy increases as the block size increases.  When it
is off-critical, entanglement entropy saturates above some critical
length which is proportional to the correlation length $\xi$ as determined in
Section \ref{subsec:cfcl}.

In Ref.~\cite{vidal2003prl}, the critical entanglement entropy is
shown to coincide with the entropy in CFT for a variety of spin chains.
Consequently, the central charge can be extracted from the critical
entanglement entropy.  As derived in~\cite{calabrese2004jsmte}, the
critical entanglement entropy satisfies
\begin{equation}
  S\approx\frac{c+\bar{c}}{6}\log\left[\frac{2L}{\pi}\sin\left(\frac{\pi L_{A}}{L}\right)\right]+k,\label{eq:eelog}
\end{equation}
for a finite lattice of total size $L$ and a sublattice of size
$L_{A}$ with periodic boundary conditions, where $c$ and $\bar{c}$ are
holomorphic and antiholomorphic central charges of the conformal field
theory and $k$ is a model-dependent constant.  For open boundary
conditions, only the holomorphic central charge is expected.

We should point out the model-dependent constant $k$ here is generally
nonzero, unlike other widely studied models.  For instance, in the
quantum Ising model, the ordered
state with zero transverse field is a product state.
The entanglement entropy of such a totally ordered state is
zero and therefore $k=0$.  On the other hand, for the two-component
Bose-Hubbard model the entanglement entropy away from the critical point
approaches that of the gapped MI system when $\Omega\rightarrow 0$. This is
vanishing only when $(U,U_{\ua\da})\rightarrow\infty$; otherwise the remaining local
particle number fluctuation contributes to the entanglement entropy as
a correction to the CFT prediction.

In extracting the central charge, there are numerical difficulties
due to the open boundary conditions.  
For this reason,
we use periodic boundary conditions.
For the XXZ Hamiltonian \eqref{eq:xxzhamiltonian}, from our numerical
calculation we confirmed the central charge $c=1/2$, corresponding
to the universality class of the transverse-field Ising model.
For the Hamiltonian \eqref{eq:hamiltonian} with
periodic boundary condition, we also successfully extracted the
central charge $c=1/2$ when $U\rightarrow\infty$,
consistent with the critical exponents we obtained above.

\section{Simulations with \MakeLowercase{i}DMRG}
\label{sec:idmrg}
\begin{figure}[t]
  \includegraphics[width=.9\linewidth]{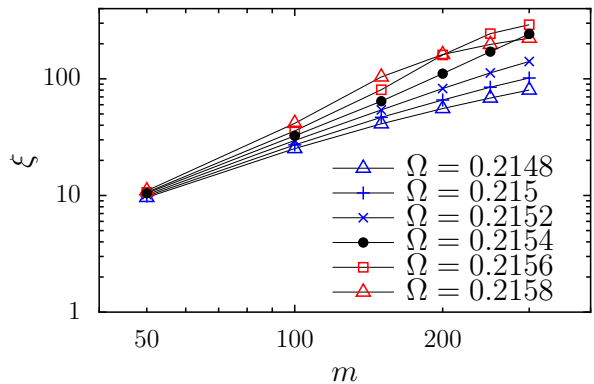}
  \caption{\label{fig:corr_leng_inf_numstat}(Color online) Correlation
    length vs. number of states for various values of the coupling
    coefficient $\Omega$ around the critical point in Mott insulator
    regime. The other parameters are $U=5, U_{\ua\da}=10$.}
\end{figure}
So far we have simulated a system only for a finite
size lattice.  The expectation values of the physical quantities are
therefore somewhat affected by finite size effects and the boundary effects
that break the translational invariance of a system.  
To remove these drawbacks, we now use the infinite DMRG~\cite{McCulloch08} that
is a better ansatz for a translational invariant system.

In Fig.~\ref{fig:mi_occ_fi}, the green crosses show the mean
occupation number imbalance obtained from iDMRG calculations.  We can
see it matches very closely the asymptotic result of the finite DMRG calculation at
the critical point.

Due to the translational invariance,
the MPS is represented by a repeated unit cell consisting of a
single site. 
While the correlation length can be extracted from the correlation function, it
can also be directly calculated from the spectrum of the transfer
matrix, which originates from exponentially decaying nature of MPS correlations
\cite{ostlund1995prl,schollwock2011ap}.

The correlation length obtained in this fashion increases as the number of
states $m$ (dimension of MPS representation) is increased.  For a gapped
noncritical system, it saturates at a certain value of $m$, while for
a gapless critical system it diverges with $m$,
and this is demonstrated in Fig.~\ref{fig:corr_leng_inf_numstat}.
Up to $m=300$, all the curves exhibit
saturation behavior except the curve for $\Omega=0.2154$.
Therefore, the critical point is close to $\Omega=0.2154$,
which agrees quite well with the value $\Omega_{c}=0.2153$ from the Binder cumulant for 
finite systems.

\begin{figure}
  \includegraphics[width=.9\linewidth]{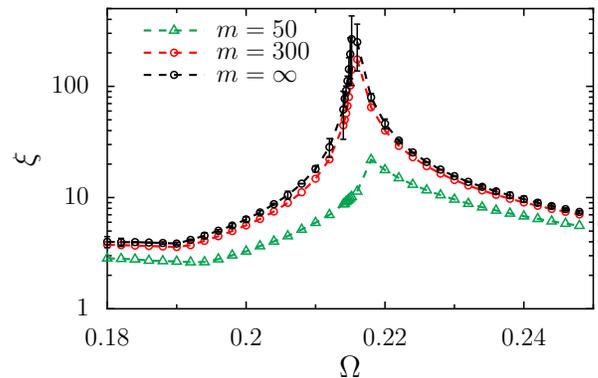}
  \caption{\label{fig:corr_leng_inf_omeg}(Color online) The
    correlation length obtained from iDMRG calculation for a
    translationally invariant infinite lattice with the increasing
    size of number of states: $m=50$ (green triangle), $m=300$ (red circle), and $m=\infty$ (black
    circle).  The data for $m=\infty$ is obtained by extrapolation. The error bar shows the error estimate in the extrapolation.
The other parameters are $U=5, U_{\ua\da}=10$.}
\end{figure}

In Fig.~\ref{fig:corr_leng_inf_omeg}, we plot the correlation length
as a function of $\Omega$ for $m=50$ and $300$.
When $m=50$, the correlation length is larger near the critical point
but the divergent behavior is not obvious.
However, it is clear for $m=300$.
We also plot the correlation length when $m$ is
extrapolated to infinity in Fig.~\ref{fig:corr_leng_inf_omeg}.
The error bars show the error increases as $\Omega$ is closer to $\Omega_{c}$.
Combining this error, we find a fit of this curve with $\xi\propto|\Omega-\Omega_{c}|^{-\nu}$
gives $\nu=0.8979\pm0.3857$ when $\Omega<\Omega_{c}$ and
$\nu=0.9621\pm0.0732$ when $\Omega>\Omega_{c}$, which is close to the known value $\nu=1$
for CFT for 1D quantum Ising model with a transverse
magnetic field, and better approximation than the fitted correlation length
for finite size calculations presented in section \ref{subsec:cfcl} above.

\section{$U(1)$-$\mathbb{Z}_{2}$ symmetry}
\label{sec:u1z2}

The total Hamiltonian \eqref{eq:hamiltonian} satisfies
$\mathbb{Z}_{2}$ symmetry because it remains unchanged if all spins
are flipped.  The ground state should preserve the same
$\mathbb{Z}_{2}$ symmetry to be symmetric or anti-symmetric, although computationally
this is not the case for symmetry breaking state when
$\Omega<\Omega_{c}$.  However, we can always unitarily transform the
Hilbert space composed of product Fock states into one composed of
symmetric and antisymmetric basis states.  The new Hilbert space can
provide us new insights into how the $\mathbb{Z}_{2}$ symmetry is
broken and restored across the critical point.

The unitary transformation for a single lattice site is,
\begin{align}
  b_{s}&=\frac{1}{\sqrt{2}}(a_{\uparrow}+a_{\downarrow})\\
  b_{a}&=\frac{1}{\sqrt{2}}(a_{\uparrow}-a_{\downarrow}).
\end{align}
 
The reversed relation can be obtained by simple linear combinations,
\begin{align}
  a_{\uparrow}&=\frac{1}{\sqrt{2}}(b_{s}+b_{a})\label{auparrow}\\
  a_{\downarrow}&=\frac{1}{\sqrt{2}}(b_{s}-b_{a}).\label{adownarrow}
\end{align}

The choice of coefficient $1/\sqrt{2}$ preserves the commutator
relation,
\begin{equation}
  [b_{s(a)},b_{s(a)}^{\dagger}]=1.
\end{equation}
Substituting \eqref{auparrow} and \eqref{adownarrow} into the three
portions of the total Hamiltonian \eqref{eq:hamiltonian}, we have the
Hamiltonian in terms of $(b_{s(a)}^{\dagger}$, $b_{s(a)})$ operators.

First the non-interacting part, Eq.~\eqref{eq:h0},
\begin{equation}
  \hat{H}_{0}  =  -J\sum_{j=1}^{L-1}\sum_{p=s,a}\left[b_{j+1,p}^{\dagger}b_{j,p}+H.c.\right].\label{u1z2h0}
\end{equation}
As no spin-flipping exists in the original Hamiltonian~\eqref{eq:h0},
symmetry is conserved during the tunneling.

The on-site interaction Hamiltonian, the first term of Eq.~\eqref{eq:hi}, between the particles of the same
component transforms to,
\begin{align}
  \hat{H}_{U}=&\frac{U}{4}\sum_{j}^{L}\sum_{p=s,a}N_{j,p}(N_{j,p}-1)\notag\\
  &+\frac{U}{4}\sum_j^{L}\left(b_{j,s}^{\dagger}b_{j,s}^{\dagger}b_{j,a}b_{j,a}+b_{j,a}^{\dagger}b_{j,a}^{\dagger}b_{j,s}b_{j,s}\right)\notag\\
  &+U\sum_{j}^{L}N_{j,s}N_{j,a},\label{eq:u1z2hu}
\end{align}
where $N_{j,p}=b_{j,p}^{\dagger}b_{j,p}$ is the number operator for
the symmetric state ($p=s$) or the antisymmetric state ($p=a$).  The
Hamiltonian describing the interaction between the two components, the second term of Eq.~\eqref{eq:hi}, becomes,
\begin{align}
  \hat{H}_{U_{\ua\da}}=&\frac{U_{\ua\da}}{4}\sum_{j=1}^{L}\Big[N_{j,s}(N_{j,s}-1)+N_{j,a}(N_{j,a}-1)\notag\\
  &-\left.\left(b_{j,s}^{\dagger}b_{j,s}^{\dagger}b_{j,a}b_{j,a}+b_{j,a}^{\dagger}b_{j,a}^{\dagger}b_{j,s}b_{j,s}\right)\right].\label{eq:u1z2hu12}
\end{align}
The two interaction Hamiltonian, Eq.~\eqref{eq:u1z2hu} and \eqref{eq:u1z2hu12}, contain terms that annihilate pairs
of symmetric bosons and create pairs of anti-symmetric bosons, and
vice versa, but with opposite sign.  


The linear coupling Hamiltonian, Eq.~\eqref{eq:hc}, between two components becomes,
\begin{equation}
  \hat{H}_{C}  =  -\Omega\sum_{j}^{L}\left(N_{j,s}-N_{j,a}\right).\label{u1z2ho}
\end{equation}
This term effectively has the function of an unbalanced chemical potential,
favoring particles in the symmetric state as $\Omega$ is increased.

\begin{figure}[t]
  \includegraphics[width=.9\linewidth]{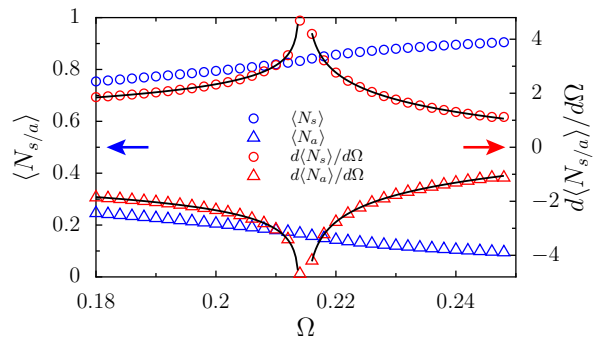}
  \caption{\label{fig:u1z2_occ_dif_mi} (Colour online) The mean
    occupation number and its derivative with respect to $\Omega$
    for symmetric states and anti-symmetric states around the critical
    point when $U=5, U_{\ua\da}=10$. The black solid curves are the fitting curves for the
    derivatives to a logarithmic function.  The arrows point to the
    corresponding $y$-axis of mean occupation number or its
    derivative.}
\end{figure}

We also employ the iDMRG algorithm to obtain the optimized iMPS with
the Hamiltonian given above.  In Fig.~\ref{fig:u1z2_occ_dif_mi}, we
plot the mean occupation number for symmetric and anti-symmetric
states.  As expected from the analysis of the linear coupling
Hamiltonian, more and more bosons occupy the symmetric states with
increasing $\Omega$.  There is a kink around the critical point
$\Omega_{c}$.  To see this kink more closely, we also plot the
derivative of the curve.  At the critical point, the derivative diverges.

We find the derivatives of the mean occupation number both for
symmetric and anti-symmetric states can be fitted with $K\log\Omega$,
as shown by the solid curves in Fig.~\ref{fig:u1z2_occ_dif_mi}.
This is further numerical evidence that
the linearly coupled two-component Bose-Hubbard model we are studying
is equivalent to a 1D quantum Ising model with transverse magnetic
field, which gives a logarithmic divergence with critical exponent $\alpha=0$~\cite{cardy1996}.
\section{Conclusion}
\label{sec:conclusion}
In this paper we have comprehensively studied the miscible-immiscible phase
transition in a linearly coupled two-component Bose-Hubbard model.  We
focus on this model in Mott insulator regime with a filling
factor one, {\it i.e.}, the total number of particles is equal to the number
of lattice sites.  We simulate this system by using both finite DMRG
and iDMRG algorithms.

We have illustrated the basic features of this phase transition. Below the
critical point $\Omega<\Omega_{c}$ the computations giving symmetry broken states show imbalanced mean
occupation number.  The imbalance decreases and disappears at the
critical point.  Above the critical point the imbalance is
zero and the $\mathbb{Z}_{2}$ symmetry of flipping the spins is
restored in the ground state.  The correlation functions of the
occupation imbalance operator show exponential decay when it is
off-critical.  The extracted correlation length diverges with
a power-law exponent $\nu$ close to one, and critical exponents $\alpha=0$ and $\beta=1/8$.

We employed the Binder cumulant to determine the critical value of linear
tunneling coefficient, and determined the phase diagram.  A
strong-coupling expansion reveals that in the Mott insulator regime this
model is equivalent to a XXZ model with a transverse magnetic field.

The phase transition is also characterized by the entanglement entropy,
which diverges logarithmically at the critical point and otherwise
saturates.  The central charge at the critical point was extracted from DMRG
calculations with periodic boundary conditions. All of these results
demonstrate conclusively that
the transition is in the universality class of the $c=1/2$ conformal field theory.

\begin{acknowledgments}
  This work has been supported by the Australian Research
  Council Centre of Excellence for Engineered Quantum Systems and the
  Discovery Projects funding scheme (Project No. DP1092513).
  After the completion of this manuscript, we became aware of other authors studying the same model~\cite{Barbiero14}.
\end{acknowledgments}

\end{document}